\documentclass[twocolumn]{aastex701}

\newcommand{\microwattsr}{\ensuremath{\mu\mathrm{W}\,\mathrm{cm}^{-2}\,\mathrm{nm}^{-1}\,\mathrm{sr}^{-1}}}

\begin{document}

\title{Correlation Between Lunar Surface and Exospheric Sodium: Effects of Albedo-Driven Temperature on Multilayer Sodium Reservoirs Rather Than Surface Abundance Variations}

\author[orcid=0000-0001-5933-058X,sname='Devaraj']{A. Devaraj}
\altaffiliation{Aryabhatta Research Institute of Observational Sciences, Devasthal Observatory, Nainital, Uttarakhand, India}
\affiliation{Department of Physics and Electronics, CHRIST (Deemed to be University), Bengaluru, India}
\email[show]{ashishdevaraj@aries.res.in}

\author[orcid=0000-0001-9199-4925,sname='Narendranath']{S. Narendranath}
\affiliation{Space Astronomy Group, U R Rao Satellite Centre, ISRO, Bengaluru, India}
\email[show]{kcshyama@ursc.gov.in} 

\author[orcid=0000-0002-7666-1062,sname='Sreeja']{Sreeja. S. Kartha}
\affiliation{Department of Physics and Electronics, CHRIST (Deemed to be University), Bengaluru, India}
\affiliation{The Centre of Excellence in Astronomy and Astrophysics (CEAA), CHRIST (Deemed to be University), Bengaluru, India}
\email[show]{\\sreeja.kartha@christuniversity.in} 

\author[orcid=0000-0001-8525-4867,sname='Pillai']{Netra S. Pillai}
\affiliation{Space Astronomy Group, U R Rao Satellite Centre, ISRO, Bengaluru, India}
\email[show]{}


\begin{abstract}
Sodium (Na) is a moderately volatile element in the lunar exosphere, released from the surface through thermal and non-thermal processes. We present a combined analysis of Chandrayaan-2 CLASS surface Na abundance, LADEE–UVS exospheric measurements, and DIVINER surface temperature data to investigate the coupling between surface and exospheric Na. Surface Na exhibits a pronounced diurnal modulation, with depletion during lunar daytime and enhancement at dawn and dusk, consistent with thermal desorption from weakly bound multilayer reservoirs ($>1$~ML). Exospheric Na shows longitudinal enhancements above low-albedo mare regions, whereas CLASS-derived surface abundances reveal no significant compositional differences between mare and highland terrains. The observed exospheric structure correlates strongly with surface temperature and albedo, implicating temperature dependent thermal desorption as the dominant release mechanism at low altitudes. These findings indicate that the spatial variability of Na release efficiency, rather than surface Na abundance, primarily governs the distribution of lunar exospheric sodium. This study places a new observational evidence on sodium retention on the lunar surface and release mechanisms and demonstrates the dominant influence of surface thermo-physics in controlling the near-surface lunar sodium exosphere.
\end{abstract}


\section{Introduction} \label{sec:introduction}
Sodium (Na) is a moderately volatile element present in the lunar Surface Boundary Exosphere (SBE). Although it's near-surface density is relatively low, about 50–100 atoms cm$^{-3}$ out of a total exospheric population of $\sim10^{5}$–$10^{6}$ atoms cm$^{-3}$, it's strong resonant scattering properties make it an excellent tracer of exospheric dynamics \citep{flynn1993,Stern1999, killen2019}. Since the pioneering ground-based detections of \cite{potter1988a, Potter1988b, Tyler1988}, numerous observational campaigns, both from the ground and from the Lunar orbit, have sought to identify the governing release processes and their relative contributions. These efforts have used both spectroscopic \citep{kuruppuaratchi2015, killen2017, killen2019, kuruppuaratchi2023} and imaging \citep{mendillo1993} techniques to study the spatial and temporal variability of Na in the SBE.

The lunar exosphere arises from a balance between multiple source and sink processes \citep{Stern1999, Stern2012, Wurz2022, Teolis2023}. A combination of thermal and non-thermal mechanisms releases surface-bound species into the SBE \citep{wurz2007, Leblanc2022, 2022SSRv..218...10W,Milillo2023}. These include thermal desorption \citep{Dukes2017}, photon-stimulated desorption (PSD) \citep{Yakshinskiy1999, sarantos2010, Devaraj2025}, solar wind sputtering \citep{smith2001, crider2003, wurz2007, milillo2011}, and micrometeorite impact vaporization (MIV) \citep{verani1998, berezhnoy2014, szalay2016, killen2019, janches2021}. The relative importance of these processes varies with local environmental conditions, such as solar illumination,and solar wind flux \citep{killen1999, wurz2007}.

Several studies indicate that the lower altitudes of the exosphere are dominated by thermal processes such as thermal desorption, which release low-energy particles that remain close to the surface \citep{killen2018understanding, Teolis2023, Milillo2023}. In contrast, higher altitudes are populated primarily through non-thermal high-energy processes—including PSD, sputtering, and MIV—which inject particles into extended ballistic trajectories or produce escape-velocity populations \citep{flynn_stern1996, potter_and_morgan1998, Leblanc2022}. This altitude dependence reflects the distinct energy distribution characteristics of each release mechanism \citep{Leblanc2022}.

The release processes described above operate on elements that are either adsorbed onto the lunar surface or chemically bound within minerals, enabling their injection into the exosphere \citep{Devaraj2025}. Consequently, the abundance of atomic species in the exosphere is often expected to mirror their distribution on the lunar surface \citep{taylor1982, johnson1991}. Early stoichiometric models predicted that major rock-forming elements such as Si, Ca, Fe, and Ti should be more abundant in the exosphere than moderately volatile elements such as Na and K \citep{flynn_stern1996, sarantos2012}. However, observational upper limits reveal that these major elements are more than an order of magnitude less abundant than predicted \citep{flynn_stern1996}. Among these species, Na remains the only element consistently detected both on the surface and in the exosphere, making it a uniquely effective tracer for investigating whether exospheric variability is tied to underlying surface composition. This forms the basis for a global correlation study of the lunar surface and exospheric Na, carried out in this work.

The Lunar Atmosphere and Dust Environment Explorer (LADEE) mission \citep{elphic2015} carried the Ultraviolet–Visible Spectrometer (UVS), which obtained orbital measurements of Na and K emissions from altitudes near 70 km \citep{Colaprete2014}. Analyses of these data by \cite{Colaprete2016} and \cite{Dawkins2022} revealed a longitudinal structure in exospheric Na, including enhancements above the mare terrains and a minimum near 0° longitude. These variations were attributed to differences in surface abundance, albedo, or regolith physical properties. However, robust global testing of these hypotheses has long been hindered by the absence of comprehensive surface Na measurements, since earlier studies relied on a small number of returned samples.

This limitation was addressed by \cite{Narendranath2022}, who produced the first global sodium abundance map using the Chandrayaan-2 Large Area Soft X-ray Spectrometer (CLASS) \citep{ pillai2021, Narendranath2025}. In the present work, we generate an updated global CLASS Na map with substantially expanded spatial coverage. To investigate potential controls on exospheric Na, we compared these new abundances with two key surface datasets: (i) DIVINER radiometer observations of surface temperature, which influence thermal desorption efficiencies, and (ii) normal albedo derived from the Lunar Orbiter Laser Altimeter (LOLA) on the Lunar Reconnaissance Orbiter, which provides a measure of reflectance independent of local topography. Together with LADEE exospheric Na measurements, these datasets enable a comprehensive assessment of the extent to which spatial variations in the lunar exosphere are governed by the underlying surface properties.

The paper is organised as follows: Section \ref{sec:data} describes the data from the CLASS instrument, LADEE, and DIVINER and LOLA. Section \ref{sec:results} presents the Analysis and Results, while Section \ref{sec:discussion} provides the discussion, and conclusion is given in Section \ref{sec:conclusion}.


\begin{figure*}
    \centering
    \includegraphics[width=2\columnwidth]{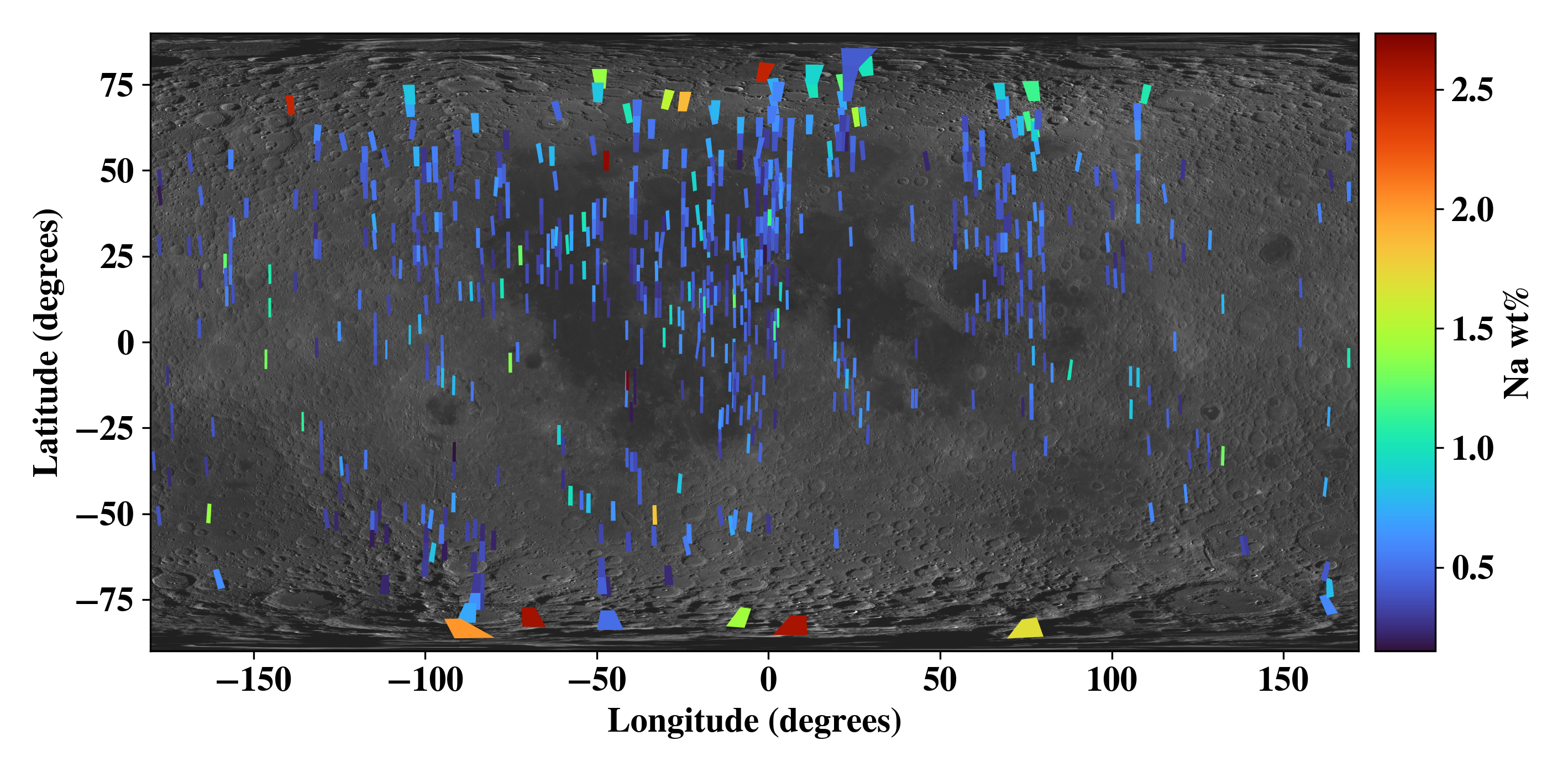}
    \caption{The Na abundance (wt\%) measured by CLASS onboard Chandrayaan-2 is shown in the figure, with ground tracks overlaid on the global albedo map from the Wide Angle Camera (WAC) of the Lunar Reconnaissance Orbiter (NASA PDS)}
    \label{fig:Na_abundance_map}
\end{figure*}

\section{Data} \label{sec:data}
\subsection{Chandrayaan-2 Large Area Soft X-ray Spectrometer (CLASS)}
CLASS is an X-Ray Fluorescence (XRF) instrument onboard the Chandrayaan-2 orbiter, which has been operating in a $\sim$100~km polar circular orbit since 2019 \citep{vanitha2020, radhakrishna2020}. CLASS comprises sixteen swept-charge devices (SCDs) that record lunar X-ray photon events on the sunlit hemisphere. The instrument operates at 0.8--20 \, keV with a spectral resolution better than 200 \, eV at 5.9 \, keV at an operating temperature of $-20^{\circ}$C. The spatial resolution is approximately $12.5$\,km $\times$ $12.5$\,km during C-class solar flares. The Level-1 data products consist of 8\,s XRF spectra from the combined SCDs, and are publicly available at \url{https://pradan.issdc.gov.in/ch2/}. The calibration and in-flight performance of CLASS are detailed in \citet{pillai2021}. Minor elements with abundances below 1\,wt\%, including Na, Cr, and Mn, are detectable during $\gtrsim$C5-class flares. It can also detect the Oxygen XRF line at 0.525 keV from the Lunar surface \citep{Devaraj2023LPICo2806.2026D}.

CLASS detects the Na~K$\alpha$ line at 1.04\,keV in the lunar XRF spectra. As this line is faint, individual 8\,s spectra are integrated in groups of twelve to obtain an effective exposure of 96\,s, which produces a ground footprint of approximately 150 km along and 12.5 km across track. To estimate the abundance of Na, we first identified observation intervals with detectable Mg, Al, and Si XRF signals and constructed integrated 96 \, s spectra for Na~K$\alpha$ analysis. Using this integration strategy, we processed all observations acquired between September 14, 2019 and May 31, 2023, to generate a global map of the abundance of Na on the lunar surface (see Figure \ref{fig:Na_abundance_map}).

The conversion of Na line flux to abundance of wt \% was performed using the \texttt{X2} abundance algorithm \citep{athiray2013a, athiray2013b}, which accounts for the incident solar spectrum, the observation geometry, and the matrix effects. Solar spectra were obtained from the XSM onboard \textit{Chandrayaan-2}; when the Sun was outside the XSM field of view, GOES-16 XRS fluxes were used to estimate the solar plasma temperature. The derived abundances, combined with \textit{Chandrayaan-2} SPICE geometry data, are used to generate maps in lunar coordinates. The analysis methodology is described in detail in \citet{2024Icar..41015898N}

\begin{figure*}[ht]
    \centering
    \includegraphics[width=0.36\textwidth]{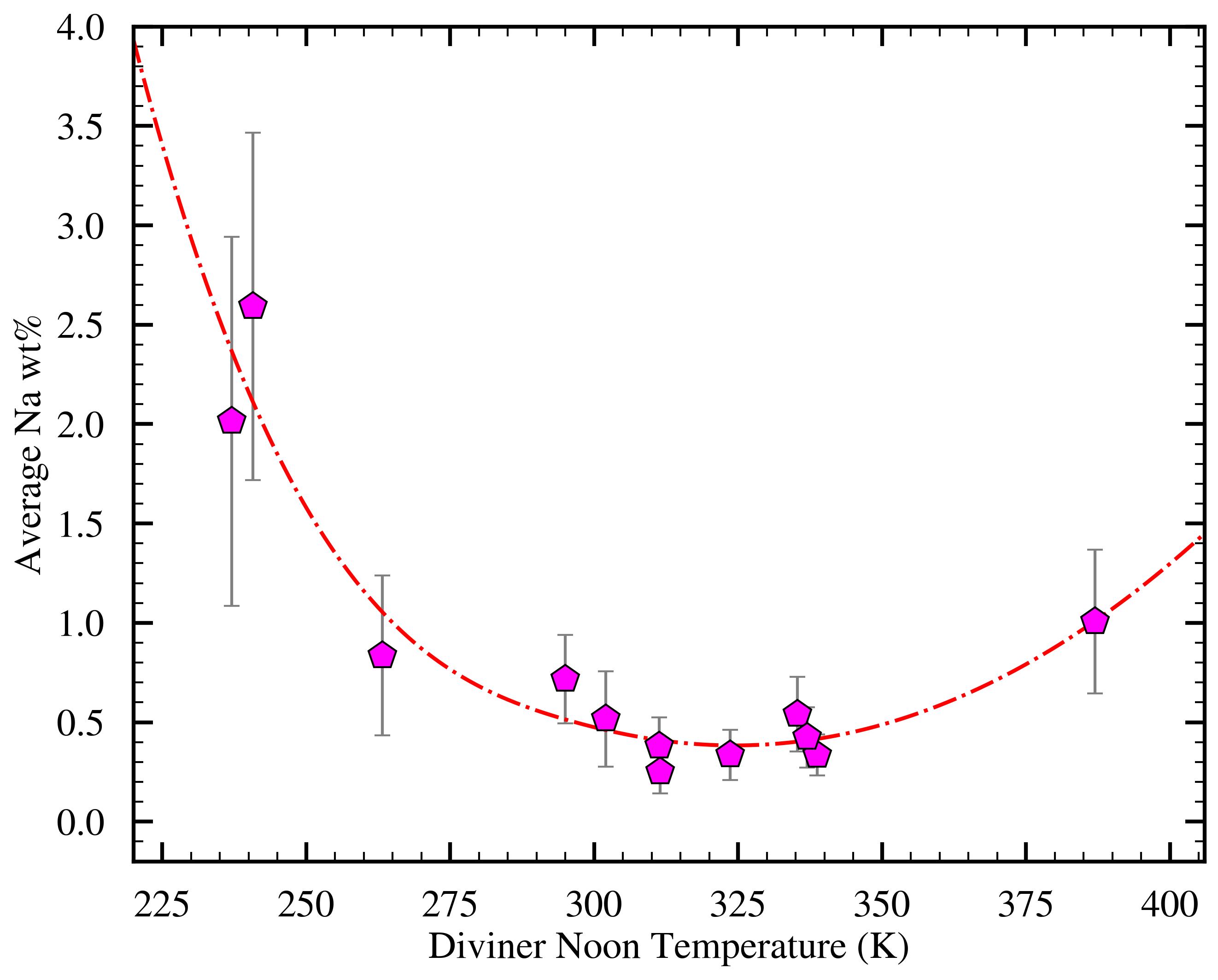}
    \hspace{0.6pt}
    \includegraphics[width=0.59\textwidth]{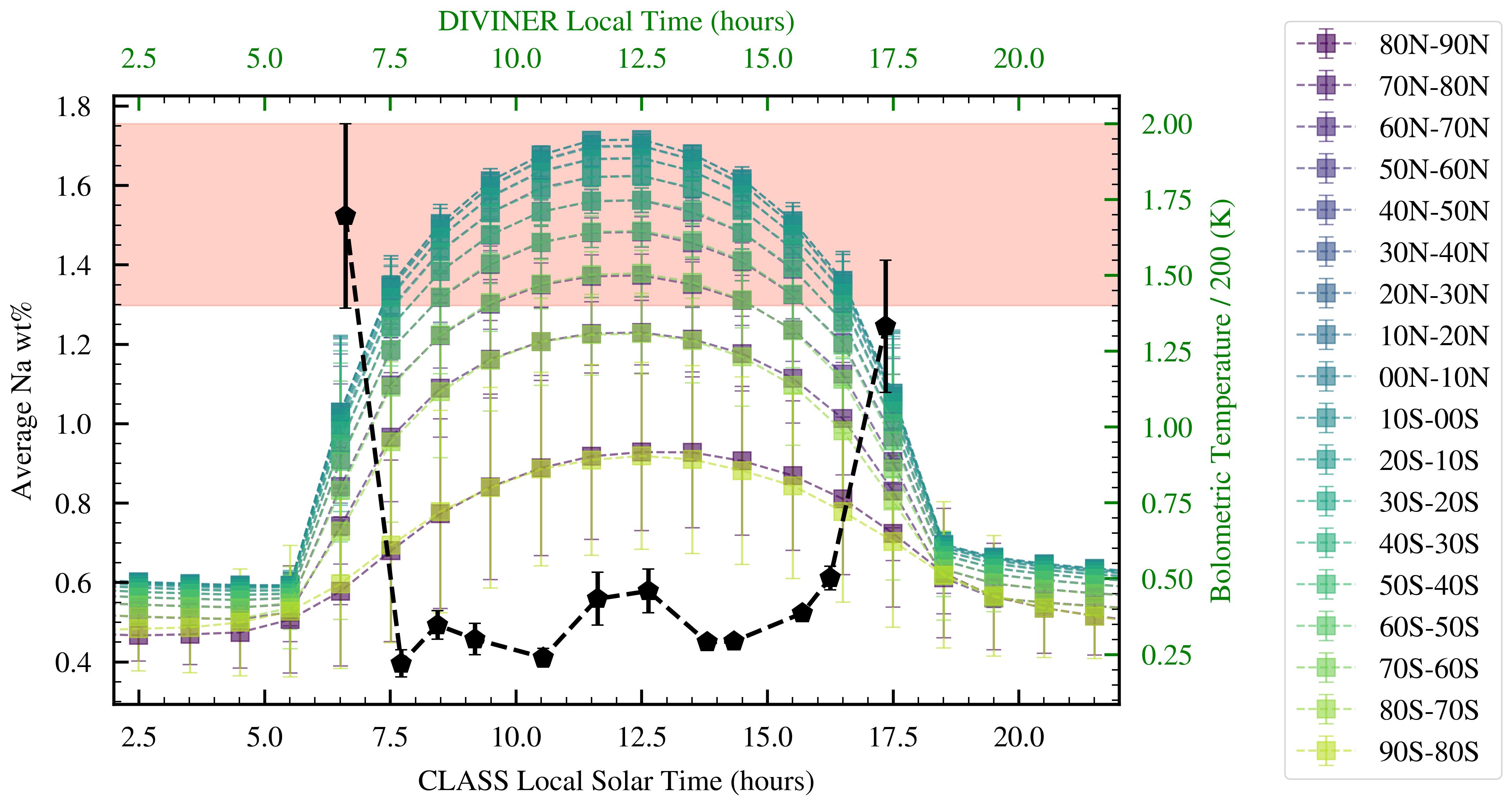}    
    \caption{The left panel shows the variation of noon-time average surface Na wt\% with the corresponding noon-time surface temperature derived from DIVINER observations. The right panel displays the Na wt\% averaged across local solar time bins (black pentagons connected with dashed line), along with the bolometric temperature measured by DIVINER as a function of local time for different latitude intervals. The red shaded region indicates the 280–400 K temperature range corresponding to the off-scale multilayer desorption peak reported in Figure 2a of \cite{Mandey1998}. This indicates the presence of multilayer Na adsorbates on Lunar surface.}
    \label{fig:naabundance_vs_lst}
\end{figure*}

\subsection{Lunar Atmosphere and Dust Environment Explorer (LADEE)}
The LADEE mission \citep{elphic2015} by NASA was launched to study the composition of the lunar exosphere and the lunar dust environment which conducted observations from low-altitude orbits around the equatorial region of the Moon. LADEE was equipped with three key scientific instruments for in situ observations: the Lunar Dust Experiment (LDEX; \citealt{horanyi2015}), the Neutral Mass Spectrometer (NMS; \citealt{mahaffy2015}), and the Ultraviolet/Visible Spectrometer (UVS; \citealt{Colaprete2014}). LDEX was designed to analyse the size and density of dust particles in the exosphere. NMS focused on detecting neutral and ionised gas species. While the UVS, operating in both Limb Mode and Occultation Mode, measured limb dust by detecting back-scattered or forward-scattered sunlight. The spectrometer covered a wavelength range of 2500-8000 $\AA$, with a spectral resolution of approximately R$\sim$900 at 5000 $\AA$, and featured a circular field of view of 0.5$^\circ$. LADEE was expected to detect various atmospheric species, including Na, K, Al, Si, Ca, Li, OH, and H$_2$, during its Limb Mode observations \citep{Colaprete2014, Colaprete2015}. Throughout the mission, LADEE monitored temporal variations in Na, K, Ar, and He \citep{elphic2015}.

The UVS data for Na measurements from LADEE were obtained from the Planetary Data System (PDS) Planetary Atmospheres Node. In this data set, the spectral line strengths of Na were derived from the UVS observations during the mid-day activities. Due to the inability to resolve the Na I D2 and D1 lines at 5889.6 and 5895.9 $\AA$, a convolution of the two lines with the instrument response function was used to derive the line strengths. This data set covers the period from September 14, 2013 to April 18, 2014, providing Na line strength measurements within a latitude range of -23.17$^\circ$ to 38.43$^\circ$ and spanning the entire longitude range from -180$^\circ$ to +180$^\circ$. The grazing altitude of the measurements made by LADEE is within $\sim$70 km.

\subsection{Lunar Reconnaissance Orbiter (LRO)}
The NASA’s LRO \citep{Chin2007, Tooley2010} was launched on 18 June 2009, and we utilize data from the Lunar Radiometer Experiment, DIVINER and the Lunar Orbiter Laser Altimeter (LOLA) for this work.  DIVINER measurements of solar reflectance and mid-infrared radiance were taken to understand the complex and extreme nature of the lunar surface thermal environment. It operates at the wavelength range of 0.3 to 400 $\mu$m \citep{Paige2010a, Paige2010b}. It produced global temperature maps with a 0.5$^\circ$ spatial resolution global dataset with a 0.25-hour local time resolution \citep{williams2017}. The DIVINER Dataset is archived at the NASA Planetary Data System Geosciences Node \citep{Paige2010a, paige2011lro}. The high-level data products (Level 2), which include gridded temperatures, were used in the analysis. The global surface temperature data represent mean bolometric temperatures for one hour of local time centered on noon, with a spatial resolution of 0.5 ppd \citep{williams2017}. The characteristics of the DIVINER instrument are described in further detail in \cite{Paige2010a}. DIVINER probes the global variations in the thermophysical properties of the regolith within the top $\sim$30 cm of the lunar regolith. Hence, the bolometric temperatures provided in this work would probe $\sim30$\,cm depth.

The LOLA onboard the LRO \citep{Smith2010_LOLA1, Smith2010_LOLA2} provides global measurements of lunar normal albedo in addition to high-precision topography. The gridded normal-albedo product used in this study is derived from LOLA Laser 1 observations obtained during the mission’s nominal phase and archived in the PDS. These data are provided at a spatial resolution of approximately 3 km per pixel at the equator in cylindrical projection. LOLA measures reflectance under zero-phase illumination, where the Sun–surface–detector geometry eliminates shadows and minimizes sensitivity to local slopes. For the low-albedo surface of the Moon, this configuration enables a robust characterization of intrinsic reflectance variations that are not influenced by small-scale topography \citep{hapke2012theory}. This makes the LOLA normal-albedo dataset a reliable reference for global photometric and compositional studies of the lunar surface. \cite{Lucey2013} provides details of the LOLA albedo calibration.

\begin{figure*}
    \centering
    \includegraphics[width=2\columnwidth]{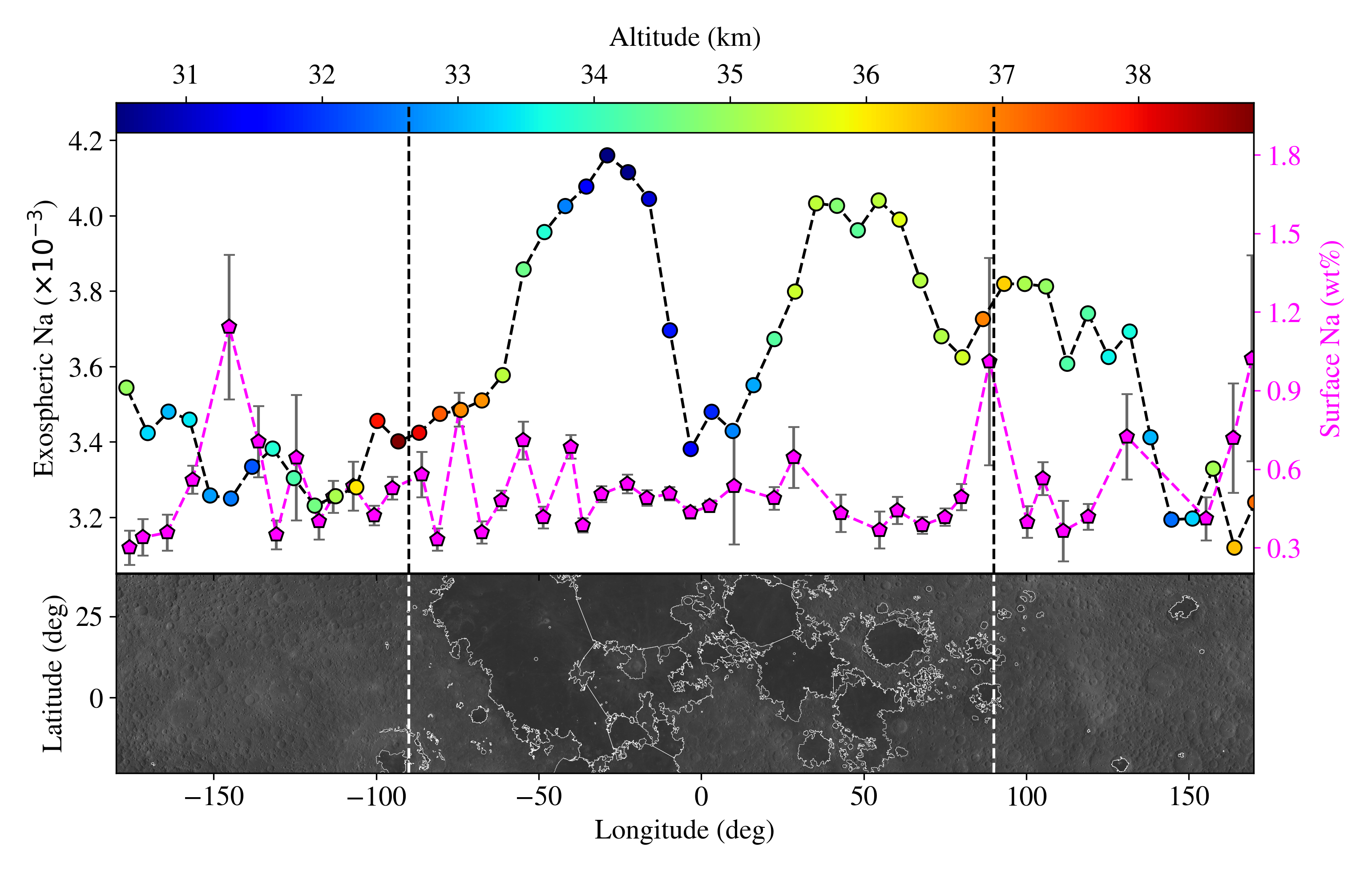}
    \caption{(Top panel) Longitudinal variation of Na exospheric line strength and surface Na abundance. Circular markers with black connected dashed line represent exospheric Na variations from LADEE, while magenta pentagon markers with the connected dashed line denote surface Na abundance (wt\%) from CLASS. The exospheric data are color-coded by the average altitude of the LADEE spacecraft within each longitude bin. (Bottom panel) LROC-WAC albedo map from the NASA PDS archive with maria boundaries overlaid. Two enhancements in exospheric Na are observed above mare regions and one above a highland region. The surface Na abundance does not detect any corresponding enhancements, indicating a no significant spatial correlation between Na distributions in the lunar exosphere and on the surface.}
    \label{fig:LADEE_longitudebin}
\end{figure*}

\section{Analysis and Results} \label{sec:results}
\subsection{Surface Na Abundance and Its Correlation with Surface Temperature}
We present a global surface Na abundance map derived from CLASS observations. The dataset presented here covers a longer observation period (up to May~2023, compared to October~2021 for the published map) and a substantially larger area of $3.17 \times 10^{6}\,\mathrm{km}^{2}$, compared to the $7.84 \times 10^{5}\,\mathrm{km}^{2}$ coverage reported by \cite{Narendranath2022}. The updated map is shown in Figure~\ref{fig:Na_abundance_map}.

To examine how the lunar environment influences surface Na, we analyzed its diurnal variation. Figure~\ref{fig:naabundance_vs_lst} (right panel) illustrates the variation of Na wt\% with local solar time (LST). Surface Na exceeds 1~wt\% before dawn and after dusk, followed by a decrease to approximately 0.4~wt\% during the lunar daytime. This reduction indicates the release of Na from the surface to the exosphere through processes such as thermal desorption and photon-stimulated desorption.

A corresponding variation is observed in the surface bolometric temperature measured by DIVINER as shown in Figure \ref{fig:naabundance_vs_lst}. Temperatures peak during the same LST interval in which the surface Na abundance reaches its minimum. The temperature range associated with the low-Na interval is approximately 280--400~K, as indicated by the red shaded region in Figure~\ref{fig:naabundance_vs_lst} (right panel). Laboratory experiments by \cite{Mandey1998}, conducted under ultrahigh vacuum conditions, reported an off-scale desorption peak ($>$1~ML) for Na from a 100~\AA\ SiO$_{2}$ film on a Re(0001) substrate within this same temperature range (see Figure 2a of \citealt{Mandey1998}). The agreement between the observed temperature range of $\sim$280--400~K, where minimum Na surface abundance is found (see Figure \ref{fig:naabundance_vs_lst}) and the laboratory desorption peak suggests that Na on the lunar surface exists in a multilayer configuration. Under such conditions, thermal desorption, a low-energy process, can become the dominant mechanism for releasing Na into the exosphere, surpassing higher-energy processes such as PSD.

\begin{figure*}
    \centering
    \includegraphics[width=\columnwidth]{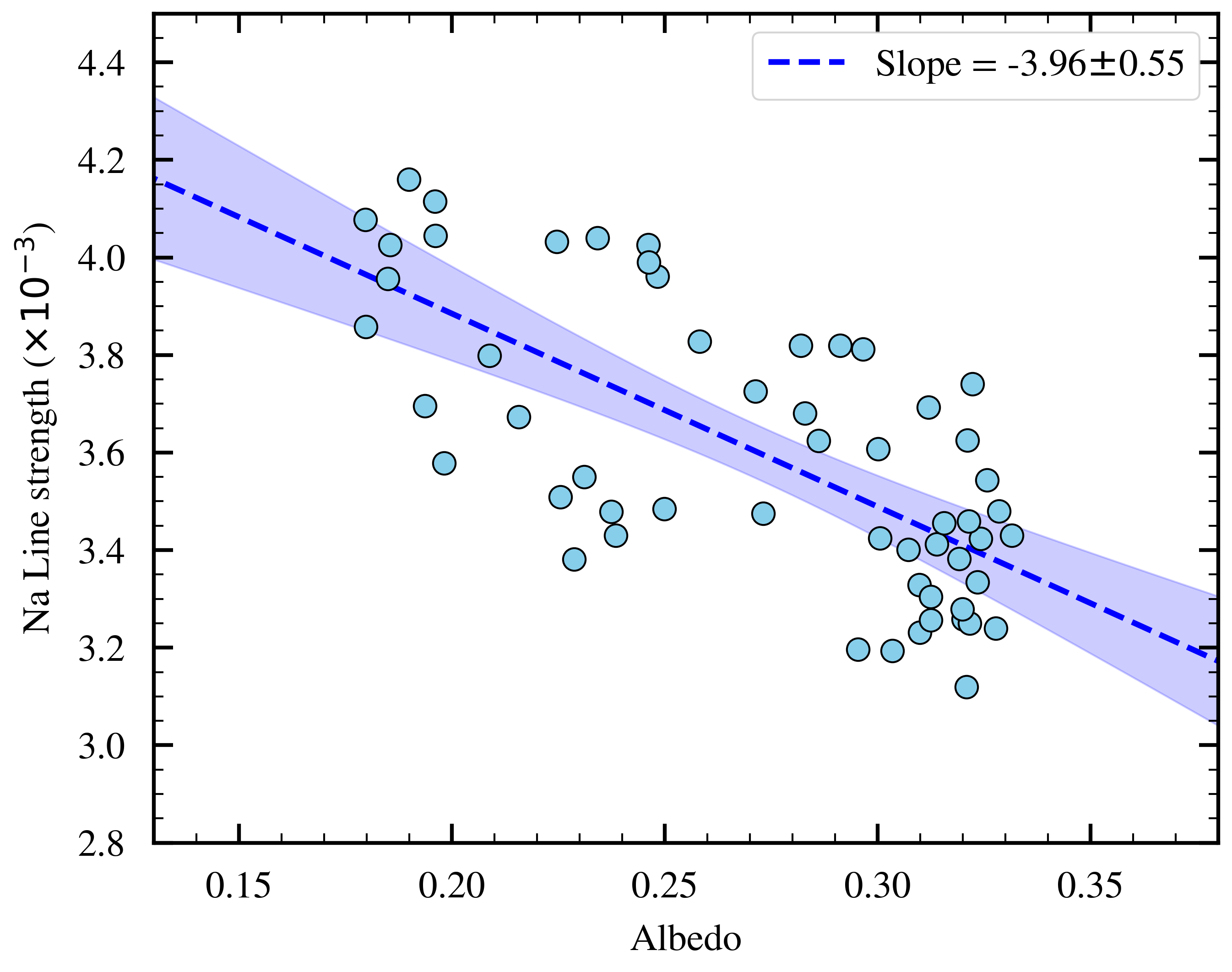}
    \hspace{1mm}
    \includegraphics[width=\columnwidth]{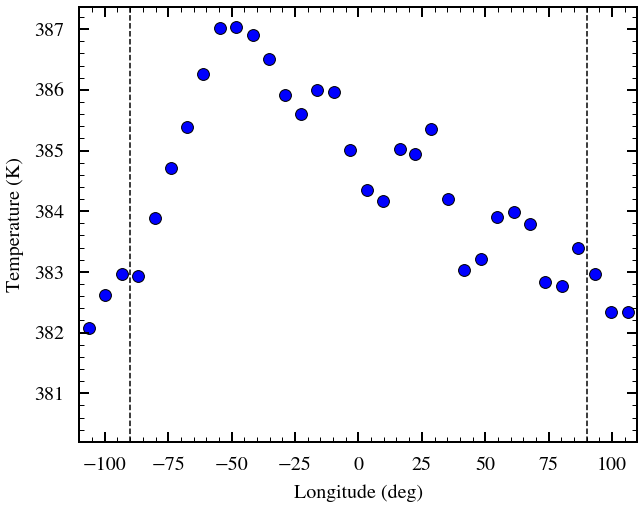}
    \caption{The left panel  depicts the correlation between Na abundance in the exosphere with the average surface albedo, indicating a negative correlation. The right panel shows the variation of average temperature from DIVINER plotted against Lunar longitude. It shows a peak above the mare region. The dashed black line shows the $\pm$90 degree longitude. This indicates that the exospheric Na is effected by the albedo driven temperature variation of the surface.}
    \label{fig:Na_exo_vs_temperature}
\end{figure*}

We also observe a correlation between noon-time surface temperature and the corresponding CLASS-derived noon-time Na abundance. The left panel of Figure~\ref{fig:naabundance_vs_lst} shows that Na abundance decreases with increasing temperature, reaching its lowest values near 320~K. This temperature again aligns with the off-scale desorption peak reported by \cite{Mandey1998}.

\subsection{Variation of Exospheric Na Concentration with Surface Temperature}
We find that the surface Na abundance decreases significantly within the temperature range of 280--400~K, consistent with thermal desorption of multilayer Na from the lunar regolith. If thermal desorption is responsible for releasing Na into the exosphere, such a decrease in surface abundance should be accompanied by a corresponding enhancement in exospheric Na concentration. To examine this connection, we analysed the Na spectral line strength measured by LADEE for the exospheric Na.

LADEE observations cover the equatorial region between latitudes $-23.17^{\circ}$ and $38.43^{\circ}$ and span the full longitude range from $-180^{\circ}$ to $+180^{\circ}$. To show the longitudinal trend, the data were binned using a longitude bin width of $6.43^{\circ}$. To avoid binning effects of the chosen bin size on the observed distribution, two independent bin-size estimation techniques were applied: (1) the Freedman--Diaconis method \citep{freedman1981}, which uses the interquartile range and is robust to non-uniform distributions, and (2) Scott's method \citep{scott1979}, which uses data standard deviation and provides a smooth density estimate for near-Gaussian distributions. The optimal number of bins returned by these methods was 52 and 56, respectively. The longitudinal distribution of exospheric Na is shown in Figure~\ref{fig:LADEE_longitudebin}. It adopt 56 bins, corresponding to $6.43^{\circ}$. Adopting 52 bins also show similar variations shown in Figure \ref{fig:LADEE_longitudebin}. The figure also includes the LRO--WAC albedo map with the lunar maria boundaries overlaid. The Na exospheric spectral line strength (in \microwattsr) exhibits clear spatial structure. Three distinct enhancements are observed: the first above Oceanus Procellarum and Mare Imbrium, with a pronounced peak between $0^{\circ}$ and $-50^{\circ}$ longitude; the second associated with multiple mare regions, peaking near $50^{\circ}$ longitude; and a third enhancement located above a highland region near $100^{\circ}$ longitude. 

The longitudinal variation of exospheric Na in Figure~\ref{fig:LADEE_longitudebin} is also color-coded by average observation altitude. Measurements at similar altitudes (\(\sim\)30--40~km) span a wide range of Na line strengths across both mare and highland regions. In particular, both high and low line strengths are observed within the same altitude range between $-50^\circ$ and $90^\circ$ longitude. Highland regions at comparable altitudes exhibit systematically lower Na line strengths relative to mare regions. The large dispersion in line strength at similar altitudes indicates the absence of a clear correlation with observation altitude, suggesting that the observed exospheric Na variability is largely independent of altitude in the alitude range considered in this dataset. In addition, the maximum uncertainty expected in Na line strength of LADEE measurements is estimated to be $\sim$8\%, obtained by combining systematic calibration ($\sim$6.5\%) and relative ($\sim$5\%) uncertainties in quadrature\footnote{\url{https://atmos.nmsu.edu/PDS/data/PDS4/LADEE/uvs_bundle/document/UVS_SIS.pdf}}.

To investigate whether the exospheric Na enhancements are influenced by surface temperature, DIVINER bolometric temperatures were filtered to the same latitude range as LADEE and binned using the identical $6.43^{\circ}$ longitude intervals. Figure~\ref{fig:Na_exo_vs_temperature} shows the resulting temperature distribution across $\pm100^{\circ}$ longitude. The temperature exhibits a peak near $-50^{\circ}$ longitude, closely matching the first peak in exospheric Na concentration. Because mare and highland terrains have different albedo, we examined the effect of albedo using LOLA reflectance. The LOLA albedo shows a negative correlation with the exospheric Na concentration: the lower-albedo mare regions exhibit higher Na line strengths, while the higher-albedo highlands show lower Na concentrations. This indicates that albedo, and therefore surface temperature, influences the release of Na into the exosphere. The correlation supports the interpretation that low-albedo mare surfaces heat more efficiently, reach higher daytime temperatures, and consequently release more Na through thermal desorption, consistent with the surface results shown in Figure~\ref{fig:naabundance_vs_lst}.

\subsection{Effect of Surface Na abundance on Exospheric Na Concentration}
Since surface composition can potentially influence exospheric composition, the Na abundance measured on the lunar surface is expected to correlate with the Na spectral line strength observed in the lunar exosphere. To test whether variations in surface Na abundance directly drive the observed exospheric Na variability, we analyzed the Na abundance map derived from CLASS within the latitude and longitude range sampled by LADEE.

The CLASS-derived surface Na abundance within the LADEE observational footprint exhibits a largely uniform distribution, with a median value of approximately 0.4~wt\% (see Figure \ref{fig:LADEE_longitudebin}). For a direct comparison with exospheric observations, the CLASS Na abundances were binned using the same $6.43^{\circ}$ longitudinal intervals employed for the LADEE data. The resulting longitudinal distribution of surface Na shows no correspondence with the exospheric Na distribution. A Pearson correlation analysis between the CLASS-derived surface Na abundance and the LADEE Na spectral line strength yields a p-value of 0.451, indicating that the correlation is not statistically significant. No systematic longitudinal variation in surface Na abundance is observed within this range. However, we observe local enhancements of Na abundance near -150$^\circ$, 90$^\circ$ and 170$^\circ$ longitudes. This enhancements show both correlation and anti-correlation with Na variations in the exosphere.

\begin{figure}
    \centering
    \includegraphics[width=\columnwidth]{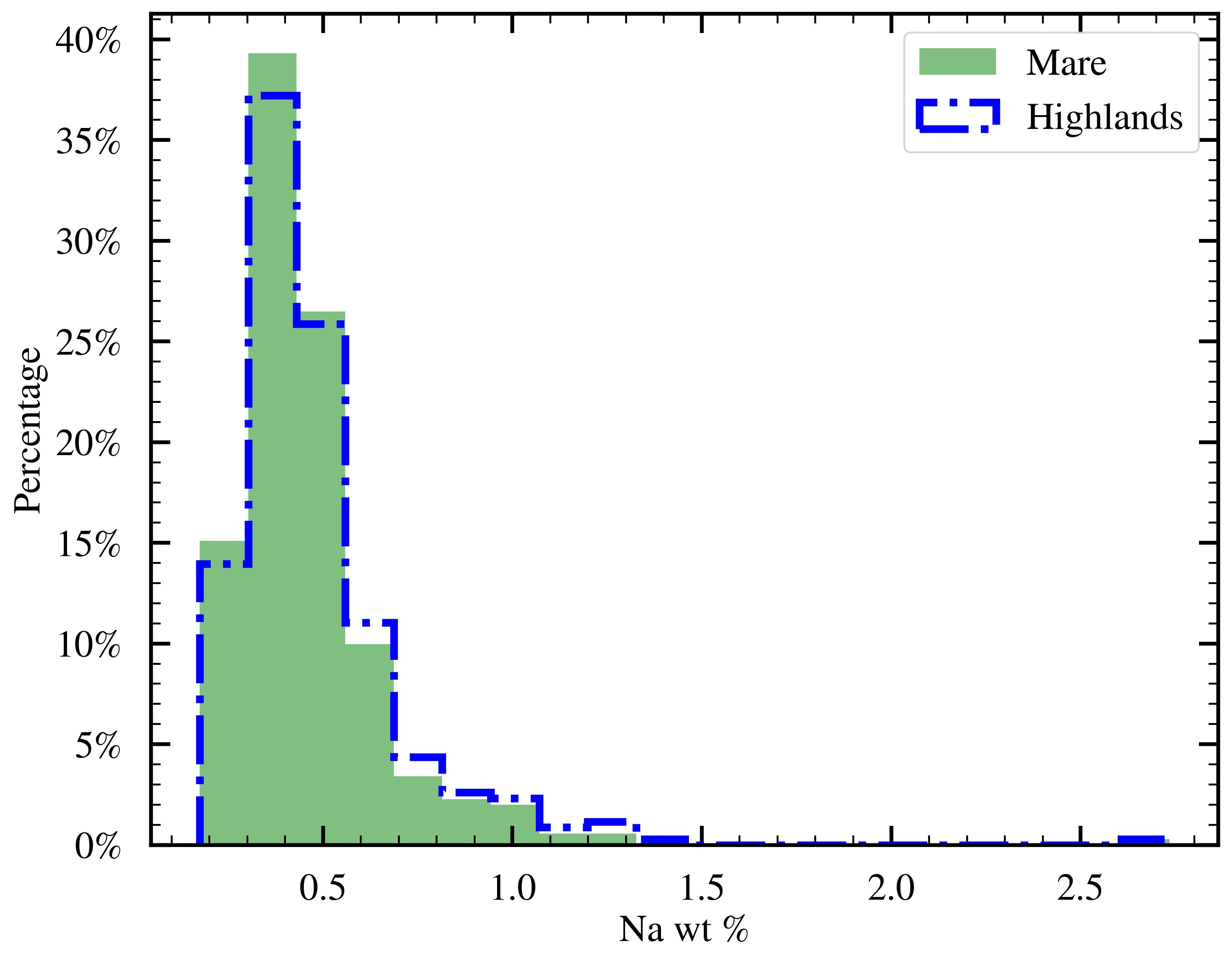}
    \caption{The percentage histogram shows the surface Na abundance (wt\%) in the mare and highland regions, as measured by CLASS. Both regions exhibit similar Na abundance.}
    \label{fig:surface_percentage_histogram}
\end{figure}

To further investigate possible compositional dichotomies on the lunar surface, the Na abundance was examined separately for mare and highland terrains. No statistically significant difference was detected between these regions: the median Na abundance is 0.41~wt\% for mare terrains and 0.42~wt\% for highlands (Figure~\ref{fig:surface_percentage_histogram}).

These results demonstrate that the spatial distribution of Na in the exosphere cannot be explained by variations in surface Na concentration as measured by CLASS. The absence of detectable surface Na variability likely could be due to both the limited sensitivity and spatial resolution of the CLASS-derived Na abundance map.

\section{Discussion} \label{sec:discussion}
\begin{figure}
    \centering
    \includegraphics[width=\columnwidth]{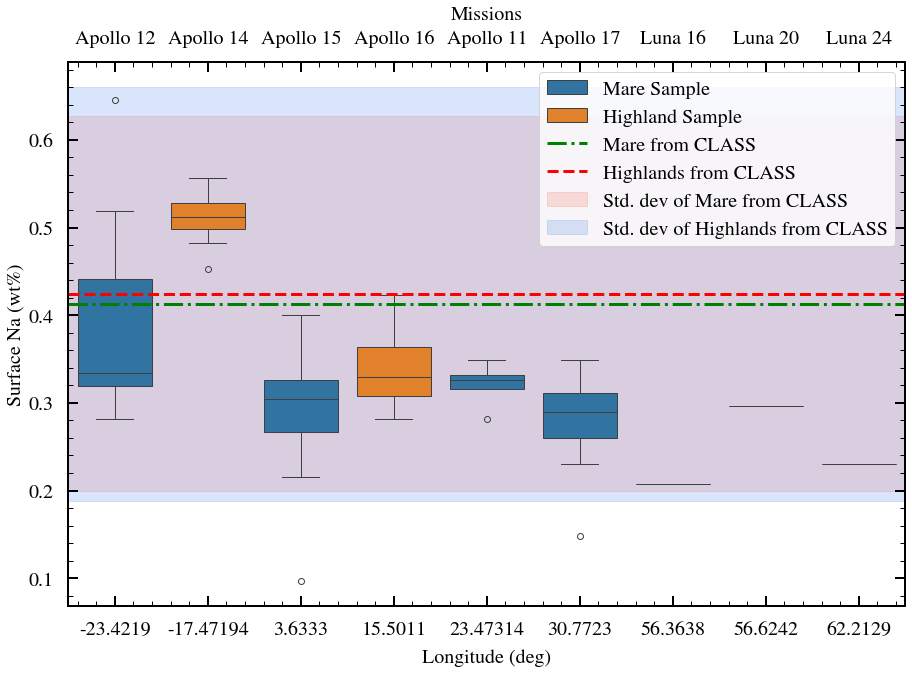}
    \caption{The comparison of ground-truth measurements from mare and highlands. The median abundance and standard deviation derived from CLASS are also indicated.}
    \label{fig:groung_truth_comparison}
\end{figure}

Sodium is a moderately volatile element that is present both on the lunar surface and in the lunar SBE. The sodium population in the lunar environment is governed by a balance between surface release processes and loss mechanisms that remove Na atoms from the Moon into interplanetary space. Consequently, variations in the lunar SBE should be intrinsically linked to the surface abundance of sodium and the physical processes responsible for its release to the exosphere. However, until recently, a global distribution of surface Na abundance measurements were unavailable. Earlier studies of lunar surface sodium were largely restricted to analyses of returned samples from various lunar missions, as shown in Figure~\ref{fig:groung_truth_comparison}.

To investigate the global behavior of sodium, the construction of a comprehensive global Na abundance map was therefore essential. Using data from the CLASS instrument onboard the Indian Chandrayaan-2 mission, the first global map of lunar surface sodium abundance was presented by \citet{Narendranath2022}. In the present work, we extend this effort by increasing the spatial coverage beyond that of the earlier map and present an updated global Na abundance map, shown in Figure~\ref{fig:Na_abundance_map}. This map serves as the basis for understanding the surface Na distribution, its role in the release processes and thereby the coupling between the lunar surface and exosphere.

\subsection{Surface sodium reservoir and release processes}
The lunar surface Na abundance exhibits a clear variation with local solar time (LST). The average Na abundance during the LST range of 6–18~hours is shown in Figure~\ref{fig:naabundance_vs_lst}. Surface Na abundance is higher during dawn and dusk, when solar-driven release agents are minimal, and decreases to $\sim$0.4~wt\% during the lunar daytime. This systematic daytime depletion indicates enhanced release of Na from the surface driven by solar illumination. While both PSD and thermal desorption are expected to operate during this interval, the close correspondence between surface Na depletion and surface temperature (see Figure \ref{fig:naabundance_vs_lst}) suggests that thermal desorption plays a dominant role at lower altitudes.

\begin{figure*}
    \centering
    \includegraphics[width=\textwidth]{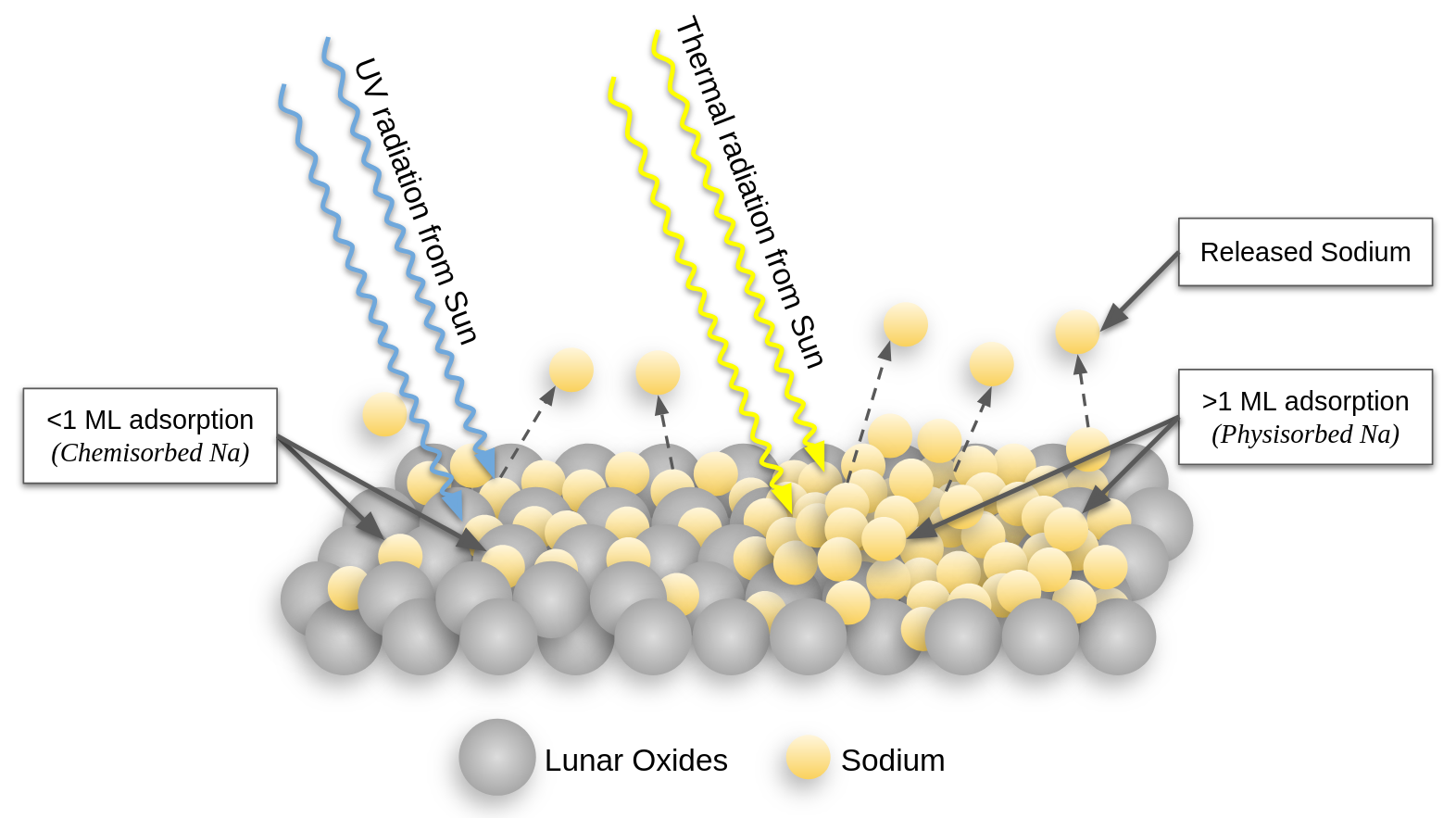}
    \caption{The figure illustrates two modes of sodium (Na) adsorption on lunar surface oxides. In the sub-monolayer (\(<1\)~ML) regime, Na is chemisorbed onto oxide surfaces through ionic bonds, which require energies \(>\)6~eV to break and are released primarily by solar UV radiation, contributing to the suprathermal exospheric population. In contrast, when available oxygen sites are depleted, Na accumulates in a multilayer (\(>1\)~ML) configuration via weak van der Waals forces (physisorption). This weakly bound Na can be released by low-energy processes such as solar-driven thermal desorption and forms the thermal exospheric population, confined to scale heights \(\lesssim\)90~km. }
    \label{fig:illustration_surface_closeup}
\end{figure*}

Laboratory studies provide important constraints on this interpretation. Experiments by \citet{Mandey1998} and \citet{yakshinskiy2000} demonstrated that the thermal desorption spectrum of Na from a 100~\AA\ thick SiO$_2$ film on a Re(0001) substrate exhibits a pronounced desorption peak within the temperature range of approximately 300–400~K. This low-temperature peak is distinct from a higher-temperature release near $\sim$700~K, corresponding to the sublimation of Na in vacuum \citep{Mandey1998, Dukes2017}. The enhanced desorption in the 300–400~K range is attributed to Na present in multilayer coverage ($>1$ monolayer, ML), which is more weakly bound to the surface than Na atoms chemisorbed on the Lunar oxides.

At sub-monolayer coverages ($<1$~ML), Na is predominantly chemisorbed on oxygen sites in surface minerals, forming strong ionic bonds with binding energies of order $\sim$6~eV \citep{Morrissey_2025}. As surface coverage approaches one monolayer, the availability of oxygen bonding sites becomes limited, suppressing further strong adsorption. In this ionic adsorption state, Na adopts a reduced effective radius of $\sim$1.4~\AA\ and remains tightly bound to the surface. Non-thermal release processes such as PSD, become effective when UV photons generate free electrons within the lattice. The surface bound Na atoms captures these free electron, increases its effective atomic radius to $\sim$2.8~\AA\ \citep{Dukes2017, Devaraj2025}, producing a repulsive interaction with the surrounding lattice that promotes desorption into the exosphere. This mechanism provides a physically consistent explanation for the preferential release of Na from strongly bound surface reservoirs under enhanced UV irradiation. This $\lesssim$1~ML configuration and the release of such Na adsorbate layer is depicted as an illustration in Figure \ref{fig:illustration_surface_closeup}.

For multilayer configurations ($>1$~ML), Na atoms are bound primarily through weaker physical interactions, such as van der Waals forces, leading to physisorption with substantially lower surface binding energies of $\sim$1–3~eV \citep{Morrissey_2025}. These weaker binding energies enable efficient release of Na through low-energy processes such as thermal desorption.

The bolometric surface temperature measured by DIVINER (Figure~\ref{fig:naabundance_vs_lst}) spans approximately 280–400~K during the lunar day, coincident with the temperature range over which the surface Na abundance decreases to $\sim$0.4~wt\%. This correspondence strongly suggests that a fraction of lunar surface Na exists as weakly bound multilayer adsorbates, in addition to Na incorporated within mineral phases. As surface temperatures rise during the lunar day, these multilayer Na reservoirs are efficiently depleted through thermal desorption, supplying Na to the lunar exosphere. This multi-layer configuration and the dominant method by which they are released to the exosphere is shown in Figure \ref{fig:illustration_surface_closeup}.

This interpretation is further supported by laboratory measurements showing that the sticking probability of Na on SiO$_2$ surfaces decreases significantly with increasing substrate temperature \citep{yakshinskiy2000}. Reduced sticking at elevated temperatures limits the re-adsorption of Na atoms onto the regolith, thereby enhancing the net release of Na into the lunar exosphere during the daytime. In addition, \citet{Narendranath2022}, using Monte Carlo simulations based on the Geant4 framework, demonstrated that the Na/Mg XRF flux ratio increases with Na adsorbate thickness. This result independently supports the presence of a multilayer Na configuration on the lunar surface. Together, the observed LST-dependent surface Na depletion and experimentally constrained adsorption and desorption behavior indicate that thermal desorption of multilayer ($>1$~ML) Na is a key driver of the diurnal variability of lunar surface sodium and its coupling to the SBE.

Sodium adsorption may vary with regolith properties such as grain size, porosity, and maturity. The enhanced surface Na abundances observed near $-150^\circ$, $90^\circ$, and $170^\circ$ longitude (see Figure \ref{fig:LADEE_longitudebin}), which exceed the global average, may reflect the dependence of Na retention on these regolith characteristics. There could be a possible concentration enhancement of Na locally due to contributions arising from micrometeorite impact-generated glasses or pyroclastic deposits.

\subsection{Coupling between surface sodium and the lunar exosphere}

The UVS onboard LADEE enabled in situ measurements of Na in the lunar exosphere, primarily sampling the near-equatorial region \citep{Dawkins2022}. These observations revealed enhanced Na emission above mare terrains. Earlier analyses by \citet{Colaprete2016} reported a symmetric enhancement near $0^{\circ}$ longitude with a local minimum at the same location, while \citet{Dawkins2022} identified a spatially non-uniform Na distribution with modest enhancements over mare regions. These studies proposed that surface composition, regolith maturity, and surface roughness—through their influence on albedo—may contribute to the observed variability.

With the availability of a global surface Na abundance map derived from Chandrayaan-2 CLASS observations, these interpretations can now be directly evaluated. In this work, we reanalysed LADEE–UVS Na line strengths using an independent longitudinal binning scheme. Our analysis reproduces the previously reported enhancement over mare regions and the local minimum near $0^{\circ}$ longitude (Figure~\ref{fig:LADEE_longitudebin}). However, comparison with the CLASS-derived surface Na abundance reveals no statistically significant difference between mare and highland terrains within the sensitivity limits of the instrument.

To quantify this result, we estimated the median surface Na abundance separately for mare and highland regions, motivated by their distinct compositional characteristics and differences in regolith maturity. The derived median abundances—0.41~wt\% for mare terrains and 0.42~wt\% for highlands—are nearly identical and consistent with ground-truth measurements from returned lunar samples (Figure~\ref{fig:groung_truth_comparison}). 

Simple stoichiometric models predict that the composition of the lunar exosphere should broadly reflect that of the lunar surface, as the exosphere is formed through the release of surface constituents into the surrounding space. Consequently, minor elements such as Na and K are expected to retain signatures of their surface abundance in the exosphere \citep{flynn_stern1996, sarantos2012}. The absence of a measurable compositional contrast in the CLASS-derived Na abundance map may be attributed to the limited sensitivity and coarse spatial resolution of CLASS, which may be insufficient to resolve localized compositional variations. Differences in the efficiency of Na release processes—such as thermal desorption, sputtering, or PSD—operating across the lunar surface are also expected to modulate the exospheric Na distribution. The observed exospheric structure therefore likely results from a combination of regional surface Na enrichment and spatial variability in surface release processes.

The energy distribution and spatial structure of the exospheric Na population are also determined by the dominant release mechanism \citep{killen2019}. Thermal desorption is a low-energy process, characterized by surface temperatures below $\sim$1000~K \citep{Dukes2017}, and preferentially produces Na atoms with low initial velocities. In a collisionless exosphere such as that of the Moon, the maximum scale-height of the exosphere with neutral Na atom depends on its initial thermal energy and lunar gravitational force on the neutral Na atom. For a surface temperature $T_{\rm surface}$, the maximum scale-height, $H$ due to thermal desorption can be approximated as

\begin{figure}
    \centering
    \includegraphics[width=\columnwidth]{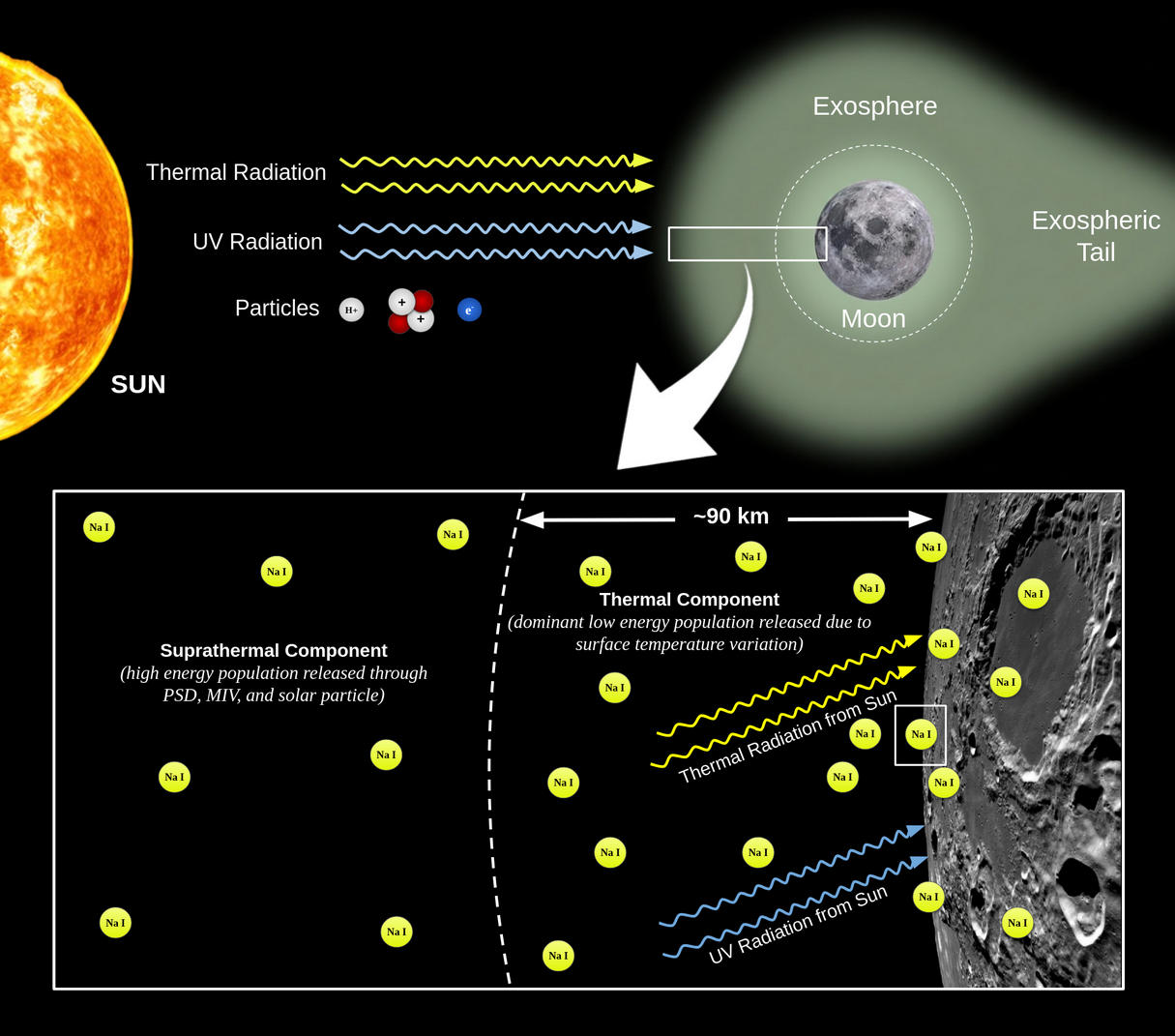}
    \caption{The top panel schematically shows solar radiation and particles driving Na release from the lunar surface and the formation of the exosphere. Ultraviolet radiation and energetic particles of solar origin impinge on the surface, releasing Na into the exosphere. The dashed white circle marks the characteristic thermal scale height (\(\sim\)90~km). Below this altitude, the exosphere is expected to be dominated by thermally desorbed Na, while higher altitudes are increasingly populated by suprathermal Na from non-thermal processes.}
    \label{fig:illustration_termal_desorption_extend}
\end{figure}

\begin{equation}
H = \frac{kT_{\rm surface}}{g_{\rm moon} m_{\rm Na}},
\label{eqn:altitude_temperature}
\end{equation}

where \(k\) is the Boltzmann constant, \(g_{\rm moon}\) is the lunar surface gravity (1.62~m~s\(^{-2}\)), and \(m_{\rm Na}\) is the mass of a Na atom (\(3.82\times10^{-26}\)~kg). For a maximum surface temperature of \(\sim\)400~K, this expression yields a maximum scale height of approximately 90~km. Solar radiation and particle fluxes, along with the corresponding extent of the exosphere produced by thermal desorption, are illustrated in Figure~\ref{fig:illustration_termal_desorption_extend}. This scale height lies well above the near-surface sampling regime of LADEE during its low-altitude equatorial orbits (see Figure \ref{fig:LADEE_longitudebin}), indicating that thermal desorption can plausibly account for a significant fraction of the Na population observed by LADEE.

Additional support for a thermally driven release mechanism is provided by the observed relationship between exospheric Na line strength and surface albedo derived from LRO–LOLA data. Surface albedo is inversely related to surface temperature, with low-albedo mare regions absorbing solar radiation more efficiently and attaining higher daytime temperatures than highland terrains. As shown in Figure~\ref{fig:Na_exo_vs_temperature}, surface temperatures peak near longitudes around $-50^{\circ}$, corresponding to Oceanus Procellarum and Mare Imbrium. These regions spatially coincide with enhanced Na line strengths observed by LADEE (Figure \ref{fig:LADEE_longitudebin}), consistent with a temperature-controlled release process.

Ground-based spectroscopic observations provide a broader physical framework for interpreting these results. Doppler line-width measurements of Na emission have revealed a two-component structure in the lunar sodium exosphere \citep{Kuruppuaratchi2018}. One component consists of a hot population with elevated Chamberlain temperatures ($T_{\mathrm{ch}}$), produced by energetic release mechanisms such as PSD, MIV, and charged-particle sputtering \citep{Ip1991, morgan_and_shemansky1991}. The second component comprises a cooler population formed through surface interactions and partial thermalization of energetic atoms \citep{Potter1988b, potter_and_morgan1991, sprague1992}. Importantly, Doppler-derived $T_{\mathrm{ch}}$ values represent the characteristic kinetic temperature of exospheric atoms rather than the physical surface temperature \citep{Mierkiewicz2014, Kuruppuaratchi2018, kuruppuaratchi2023}. Higher-energy processes such as PSD, solar wind ion sputtering, and MIV contribute to the extended exosphere by imparting higher energies to the released species. Among these, PSD is generally the dominant contributor, while MIV becomes significant episodically during periods of enhanced meteoroid fluxes \citep{verani1998, sarantos2010, szalay2016, janches2021, berezhnoi2023}. Solar wind ion sputtering is comparatively less efficient than PSD \citep{szabo2018, jaggi2024}. In contrast, thermal desorption is largely confined to near-surface altitudes, producing scale heights below $\sim$90~km and characteristic temperatures $\lesssim$400~K \citep{sprague1992}. Consequently, the Na population sampled by LADEE is expected to be dominated by the cooler, thermally desorbed component according to the broader physical framework established by previous ground based observations.

Taken together, we present an updated global framework of the configuration of the Na adsorption layer on the lunar surface and its coupling to the lunar exosphere by combining Chandrayaan-2 CLASS surface abundance measurements with LADEE--UVS exospheric observations and Diviner and LOLA surface temperature and albedo data, along with comparisons to laboratory experiments.

Surface Na exhibits pronounced diurnal modulation, with depletion during lunar daytime and enhancement at pre-dawn and post-dusk, closely tracking Diviner-measured surface temperatures. Minimum surface Na abundances occur within 280--400~K, consistent with laboratory measurements of enhanced thermal desorption from multilayer ($>1$~ML) Na reservoirs. This indicates that weakly bound ($\sim$1--3~eV) multilayer Na dominates the surface adsorbate inventory. During lunar daytime, when surface temperatures reach $\sim$280--400~K, thermally activated release from weakly bound multilayer ($>1$~ML) Na efficiently supplies the near-surface exosphere at altitudes $\lesssim$100~km. PSD, while continuously active under solar UV irradiation, primarily affects more strongly bound Na surface populations ($\lesssim$1~ML) and Na incorporated within mineral phases, producing higher-energy ejecta capable of populating larger scale heights \citep{Devaraj2025}. Consequently, PSD contributes predominantly to the extended exosphere, whereas thermal desorption governs near-surface enhancements and diurnal variability.

Our independent analysis of LADEE--UVS data confirms the longitudinal enhancements in exospheric Na over mare regions reported by \cite{Colaprete2016} and \cite{Dawkins2022}. However, comparison with CLASS-derived surface Na abundances reveals no statistically significant compositional difference between mare and highland terrains, indicating that spatial compositional variations play a secondary role in the observed exospheric Na variability. Instead, exospheric Na shows a strong dependence on surface temperature and albedo, implicating thermally driven release as the dominant mechanism at lower altitudes.

Overall, the lack of correlation between exospheric Na line strength and surface Na abundance, together with the strong dependence on surface temperature and albedo, demonstrates that Na release efficiency---rather than absolute surface compositional variations---primarily controls the spatial and temporal structure of the lunar sodium exosphere.

\section{Conclusion} \label{sec:conclusion}
We present a framework for a global lunar surface--exosphere coupling by integrating multiple observational datasets and laboratory experiments for the first time. The key conclusions are:

\begin{itemize}

\item Observational evidence supports the presence of a dominant multilayer ($>1$~ML) Na adsorbate reservoir on the lunar surface. Under these conditions, thermal desorption becomes more efficient than higher-energy processes such as PSD at low altitudes.

\item Exospheric Na exhibits enhancements over mare regions; however, no statistically significant correlation is found with surface Na abundance. Comparison with low-spatial-resolution CLASS-derived Na abundances reveals no compositional distinction between mare and highland terrains within detection limits, indicating that exospheric variability is governed primarily by release efficiency rather than intrinsic surface abundance.

\item Surface Na abundance shows a clear dependence on LST, with reduced abundances during daytime (280--400~K), consistent with temperature-driven thermal desorption, indicating thermal desorption is the dominant process at low altitudes.

\end{itemize}

Collectively, these results demonstrate that the efficiency of Na release mechanisms--rather than absolute surface abundance--controls the spatial and temporal structure of the lunar sodium exosphere, and highlight the key role of weakly bound multilayer reservoirs in governing surface--exosphere coupling at low altitudes.

\begin{acknowledgments}
We thank the reviewer for their careful evaluation of the manuscript and for their constructive comments, which have helped improve its clarity. This research was funded by the Indian Space Research Organisation (ISRO) under the Chandrayaan-2 Announcement of Opportunity (AO) project (Sanction No. DS\_2B-13013(2)/8/2022-Sec.2). We also acknowledge facility support from the Department of Science and Technology (DST), Government of India, under the FIST programme (SR/FST/PS-I/2022/208). We further acknowledge the research and institutional support provided by CHRIST (Deemed to be University).
\end{acknowledgments}

\facilities{Chandrayaan-2 (CLASS), LRO (DIVINER and LOLA), LADEE (UVS)}

\bibliography{references}{}
\bibliographystyle{aasjournalv7}

\end{document}